\documentclass{aa}  

\bibpunct{(}{)}{;}{a}{}{,} 

\usepackage{graphicx}
\usepackage{txfonts}
\begin{document}

   \title{Combining high-dispersion spectroscopy (HDS) with high contrast imaging (HCI): Probing rocky planets around our nearest neighbors}

\authorrunning{I. Snellen et al.}
\titlerunning{HDS+HCI: Probing rocky planets around our nearest neighbours}
   \author{I. Snellen
          \inst{1},
           R. de Kok
          \inst{1,2},
          J.L. Birkby
               \inst{1,3,4},
          B. Brandl
          \inst{1},
          M. Brogi
          \inst{1,5,6},
          C. Keller
          \inst{1},
          M. Kenworthy
          \inst{1},
          H. Schwarz
          \inst{1},
          R. Stuik
          \inst{1}
          } 
          
   \institute{Leiden Observatory, Leiden University, Postbus 9513, 2300 RA, Leiden, The Netherlands\\
              \email{snellen@strw.leidenuniv.nl}
   \and
   SRON, Netherlands Institute for Space Research, Sorbonnelaan 2, 3584 CA Utrecht, The Netherlands
   \and
   Harvard-Smithsonian Center for Astrophysics, 60 Garden Street, Cambridge MA 02138, USA
   \and
   NASA Sagan Fellow
   \and
   Department of Astrophysical and Planetary Sciences, University of
   Colorado, Boulder CO 80309, USA
  \and
   NASA Hubble Fellow
          }

   \date{}

\abstract{Ground-based high-dispersion (R$\sim$100,000) spectroscopy (HDS) is proving to be a powerful technique with which to characterize extrasolar planets. The planet signal is distilled from the bright starlight, combining spectral and time-differential filtering techniques. In parallel, high-contrast imaging (HCI) is developing rapidly, aimed at spatially separating the planet from the star. While HDS is limited by the overwhelming noise from the host star, HCI is limited by residual quasi-static speckles. Both techniques currently reach planet-star contrast limits down to $\sim 10^{-5}$, albeit for very different types of planetary systems.}
{In this work, we discuss a way to combine HDS and HCI (HDS+HCI). For a planet located at a resolvable angular distance from its host star, the starlight can be reduced up to several orders of magnitude using adaptive optics and/or coronography. In addition, the remaining starlight can be filtered out using high-dispersion spectroscopy, utilizing the significantly different (or Doppler shifted) high-dispersion spectra of the planet and star. In this way, HDS+HCI can {\sl in principle} reach contrast limits of $\sim 10^{-5} \times 10^{-5}$, although in practice this will be limited by photon noise and/or sky-background. In contrast to current direct imaging techniques, such as Angular Differential Imaging and Spectral Differential Imaging, it will work well at small working angles and is much less sensitive to speckle noise. For the discovery of previously unknown planets HDS+HCI requires a high-contrast adaptive optics system combined with a high-dispersion R$\sim$100,000 integral field spectrograph (IFS). This combination currently does not exist, but is planned for the European Extremely Large Telescope.}
{We present simulations of HDS+HCI observations with the E-ELT, both probing thermal emission from a planet at infrared wavelengths, and starlight reflected off a planet atmosphere at optical wavelengths. For the infrared simulations we use the baseline parameters of the E-ELT and METIS instrument, with the latter combining extreme adaptive optics with an R=100,000 IFS. We include realistic models of the adaptive optics performance and atmospheric transmission and emission. For the optical simulation we also assume R=100,000 IFS with adaptive optics capabilities at the E-ELT.} 
{One night of HDS+HCI observations with the E-ELT at 4.8 $\mu$m ($\Delta \lambda = 0.07 \mu$m) can detect a planet orbiting $\alpha$\,Cen\,A with a radius of R=1.5 R$_{\rm{earth}}$ and a twin-Earth thermal spectrum of T$_{\rm{eq}}$=300 K at a signal-to-noise (S/N) of 5. In the optical, with a Strehl ratio performance of 0.3, reflected light from an Earth-size planet in the habitable zone of Proxima Centauri can be detected at a S/N of 10 in the same time frame. Recently, first HDS+HCI observations have shown the potential of this technique by determining the spin-rotation of the young massive exoplanet $\beta$ Pictoris b.}
{The exploration of the planetary systems of our neighbor stars is of great scientific and philosophical value. The HDS+HCI technique has the potential to detect and characterize temperate rocky planets in their habitable zones. Exoplanet scientists should not shy away from claiming a significant fraction of the future ELTs to make such observations possible.}

\keywords{Infrared: planetary systems – Methods: data analysis – Techniques: imaging spectroscopy - Techniques: high angular resolution}

   \maketitle
%

\section{Introduction}

Since the late 1990s it  has been recognized that high-dispersion spectroscopy (HDS) could be a powerful way to characterize extrasolar planet atmospheres. First attempts to use HDS were focussed on detecting optical starlight reflected off hot Jupiter atmospheres. However, their optical albedos turned out to be low (e.g. \citealp{Charbonneau99}; \citealp{Colliercameron04}; \citealp{Leigh03a}; \citealp{Leigh03b}), making these measurements very challenging. Early attempts to detect molecular absorption features in the thermal emission spectra of hot Jupiters with HDS also resulted in upper limits (\citealp{Wiedemann01}; \citealp{Deming05}).

Recently, the availability of more stable infrared spectrographs, in particular the CRyogenic high-resolution InfraRed Echelle Spectrograph (CRIRES; \citealp{Kaeufl04}) on the Very Large Telescope (VLT, European Southern Observatory), in combination with new observational strategies, has led to the first detections. \citet{Snellen10} measured carbon monoxide at 2.3 $\mu$m in the transmission spectrum of HD 209458b, revealing the orbital motion of the planet.  \citet{Brogi12} found the same molecular species in the thermal spectrum of the non-transiting planet $\tau$\,B\"ootis\,b, determining its mass and orbital inclination (see also \citealp{Rodler12}), and it was also detected in the thermal spectra of HD 189733\,b (\citealt{deKok13}; \citealt{Rodler13}) and tentatively 51\,Peg\,b \citep{Brogi13}. Recently, \citet{Birkby13} measured water absorption in the thermal spectrum of hot Jupiter HD\,189733\,b at 3.2 $\mu$m using HDS with CRIRES, and \citet{Lockwood14} in that of  $\tau$\,B\"ootis\,b using the Keck telescope.

In parallel to HDS, high-contrast imaging (HCI) is rapidly developing, driven by improvements in adaptive optics technology and coronographic concepts. This technique is now well established for young, massive self-luminous planets  at significant orbital distances from their host star (e.g. \citealp{Chauvin05}, \citealp{Lagrange09}, \citealp{Marois08}, \citealp{Kalas08}). Several new HCI facilities are coming online, such as GPI on the Gemini Telescope (\citealp{Macintosh14}), SPHERE on the Very Large Telescope (\citealp{Beuzit10}), and SCExAO on Subaru (\citealp{Jovanovic13}), which will push these observations to closer and cooler planets. 

In this paper we discuss a way to combine high dispersion spectroscopy with high-contrast imaging (HCI). For a planet located at a resolvable angular distance from its host star, the starlight can be reduced up to several orders of magnitude using adaptive optics and/or coronography. In addition, the remaining starlight can be filtered out using high-dispersion spectroscopy, utilizing the significantly different (or Doppler shifted) high-dispersion spectra of the planet and star. In Sect. 2 we give technical descriptions for HDS, HCI, and HDS+HCI. In Sect. 3 we describe the instrument requirements for HDS+HCI,
which basically come down to a high-contrast adaptive optics system combined with a high-dispersion R$\sim$100,000 integral field spectrograph (IFS). This combination currently does not exist, but is planned for the European Extremely Large Telescope. In Sect. 4 we describe simulations of infrared HDS+HCI observations with the METIS instrument on the European Extremely Large Telescope (E-ELT), and in Sect. 5 that of optical HDS+HCI observations with a hypothetical instrument on the same telescope. The results are discussed in Sect. 6. 

\section{Technical descriptions}
\subsection{High dispersion spectroscopy (HDS)}

High dispersion spectroscopy makes use of the fact that at high spectral dispersion molecular bands resolve into tens to hundreds of individual lines, which shift in wavelength during observations thanks to a change in the radial component of the orbital motion of the planet. For hot Jupiters, which have orbital velocities of $>$100 km sec$^{-1}$, this change in wavelength can be up to a several times the spectral resolution per hour. This is used to separate the planet spectrum from the quasi-stationary telluric and stellar components which strongly dominate the observed spectra. The measurement of a molecular species is subsequently enhanced by cross-correlating the observed spectra with a model template spectrum, such that the signals from the individual lines are optimally combined (e.g. \citealt{Brogi13}). 

Neglecting the interference by telluric absorption, the achieved signal-to-noise ratio (S/N) of HDS is to first order given by
\begin{equation}
\rm{S/N} = \frac{S_{\rm{planet}}}{\sqrt{S_{\rm{star}}+\sigma_{\rm{bg}}^2+\sigma_{\rm{RN}}^2+\sigma_{\rm{Dark}}^2}} \sqrt{N_{\rm{lines}}},
\end{equation}
where $S_{\rm{planet}}$ is the planet signal, $S_{\rm{star}}$ is the signal from the star (both in units of photons per resolution element) , and $\sigma_{\rm{bg}}$ , $\sigma_{\rm{RN}}$, and $\sigma_{\rm{Dark}}$  are the photon shot noise from the sky and telescope background, the readout noise, and the noise from the Dark current respectively. $N_{\rm{lines}}$ is a multiplication factor that takes into account the number and strength of the individual planet lines targeted. For most HDS observations, $\sigma_{\rm{bg}}$, $\sigma_{\rm{RN}}$ and $\sigma_{\rm{Dark}}$ can be neglected because the stars are bright, and the backgrounds low. Combining the thousands of lines in the optical spectrum of a sunlike star typically results in a S/N enhancement of a factor of $\sim$30 ($N_{\rm{lines}} \sim$ 1000, \citealp{Colliercameron04}). In this way, \citet{Leigh03a} reached a 1$\sigma$ planet/star contrast limit of 1.3$\times 10^{-5}$ for the HD75289b system (V=6.3)  by combining 684 spectra taken over 4 nights with UVES on the VLT. For $\tau$ B\"ootis b (K=3.36) at 2.3 $\mu$m, \citet{Brogi12} reached a 1$\sigma$ CO line-contrast of $\sim$2$\times$10$^{-5}$  with respect to the stellar spectrum in 15 hours of CRIRES VLT observations, utilizing a factor of $\sim$5 enhancement by combining the signal from two-dozen CO lines. \citet{Birkby13} reached a 1$\sigma$ H$_2$O line contrast of $<$3$\times$10$^{-4}$ in 5 hours of observations at 3.2$\mu$m for HD189733 b (K=5.54) using the same instrument. Because of the strong telluric absorption in this wavelength regime, the S/N enhancement from combining numerous but weak water lines is only a factor of $\sim$3.

\citet{deKok14} identified new opportunities of using CRIRES to probe hot Jupiters, showing that several other wavelength regions promise to give strong molecular signals, such as around 3.5$\mu$m for dayside spectroscopy and at 1.6 $\mu$m for transmission spectroscopy. In addition, it is expected that in certain cases the planet night-side may even produce stronger signals than its dayside. Although the nightside temperature is lower and therefore the planet signal fainter, the absorption lines themselves may be stronger as a result of a steeper atmospheric temperature-pressure profile. 

\citet{Snellen13} explored future possibilities of high-dispersion {\sl transmission} spectroscopy to detect molecular oxygen in nearby Earth-like exoplanets (see also \citealp{Rodler14}). They showed that the transmission signal of the oxygen A-band from an Earth-twin orbiting a small red dwarf star is only a factor of 3 smaller than that of carbon monoxide detected by \citet{Brogi12} in the hot Jupiter tau Bo\"otis b, albeit such a star will be orders of magnitude fainter. If Earth-like planets are very common, the planned extremely large telescopes could detect an oxygen signal within a few dozen transits, but this could take decade(s), since only a few transits would be favorably observable from one location on Earth per year. Detection of an oxygen signal from an Earth-like planet in the habitable zone of  a solar type star could take centuries,  since such transits are so rare. We note that this same idea, HDS of the oxygen A band, was already suggested more than a decade ago by \citet{Webb01}, but they argued it could take 10 hours or less on an 8m telescope.  

\subsection{High-contrast imaging (HCI)}

 Planet detections with HCI lead to planet characterization when observed at multiple wavelenths. Broadband photometry have lead to estimates of planet effective temperatures and radii, and possible prevalance of clouds (e.g. \citealp{Currie11}, \citealp{Currie13}, \citealp{Bonnefoy13}). 
However, for ground based observations, the turbulent atmosphere degrades the
diffraction limited images of stars into a seeing limited disk. All
large telescopes use an adaptive optics system to measure the
atmospheric distortion with a wavefront sensing (WFS) camera and then
apply optical correction with a deformable mirror to drive the sensed
wavefront aberrations to a null. The resultant diffracted halo of
light from the primary star dominates the noise contribution at the
location of the exoplanet (see middle panel of Figure 1). In addition, the optical
WFS camera path aberrations and science instrument optical path aberrations vary slowly with time,
changing the relative intensity of the diffraction halo at a level of
10$^{-4}$ to 10$^{-5}$ of the stellar peak flux. Differencing two images taken
several minutes apart shows the presence of quasi-static speckles
similar in angular size to potential faint companions. These are attributed to
the time-varying nature of the non-common path aberrations between the
adaptive optics system WFS camera and science camera. 
 Several high contrast observing techniques and post-processing algorithms have been developed to form a two-step process to estimate the science PSF as a
function of time during the science exposures that gain extra orders of magnitude in planet/star contrast. These include Angular Differential Imaging (ADI; \citealp{Marois06}), Spectral Differential Imaging (SDI; \citealp{Racine99}), Locally Optimized Combinations of Images (LOCI; \citealp{Lafreniere07}), and Principal Component Analysis (PCA; \citealp{Amara13}, \citealp{Soummer12}, \citealp{Meshkat14}). Some of these techniques can also be used in combination to get the best contrast.

Coronagraphs are angular filters that reject on-axis light whilst
transmitting light from nearby off-axis faint companions, reducing the
diffraction halo from the primary star through optics introduced in
the focal plane and pupil planes of the science camera (see \citealp{Guyon06a}
for a review). Coronagraphs reduce the flux in diffraction structures
and so the effects of non-common path errors are correspondingly
reduced. Several coronagraphic designs for space-based high contrast
designs (1e-9 to 1e-10) have been developed (Phase-induced amplitude apodization or PIAA - \citealp{Guyon06b}; Vortex
Coronaraph, \citealp{Mawet09}; Band Limited Coronagraph, \citealp{Vanderbei03}; Cash 2006) but all are sensitive to pointing errors and vibrations that the
AO systems cannot remove, and are chromatically limited. For ground based telescopes, the raw
constrast is limited to 10$^{-5}$ at a few $\lambda$/D because of the time delay
between the WFS and the AO system response, (\citealp{Guyon12}). In these
cases, pupil plane coronagraphs that use phase apdization (APP;
\citealp{Kenworthy07}, \citealp{Codona04}; broadband APP \citealp{Otten14}) or amplitude apodization (\citealp{Carlotti13}) can provide robust suppression at small angular separations (\citealp{Kenworthy13}; \citealp{Meshkat14}) in science
instruments without a dedicated coronagraphic optical design. An advanced APP with a limited dark area in the shape of a slot can be designed with a contrast of 10$^{-10}$ or better with a transmission of 95\% (Keller et al. in preparation). When manufactured as a vector APP (Snik et al. 2012), such coronagraph will work over a wavelength range of an octave while being largely immune to any remaining tip-tilt errors in the incoming wavefront.

\subsection{Combining HDS and HCI}

\begin{figure*}
\centering
\includegraphics[width=1.0 \textwidth]{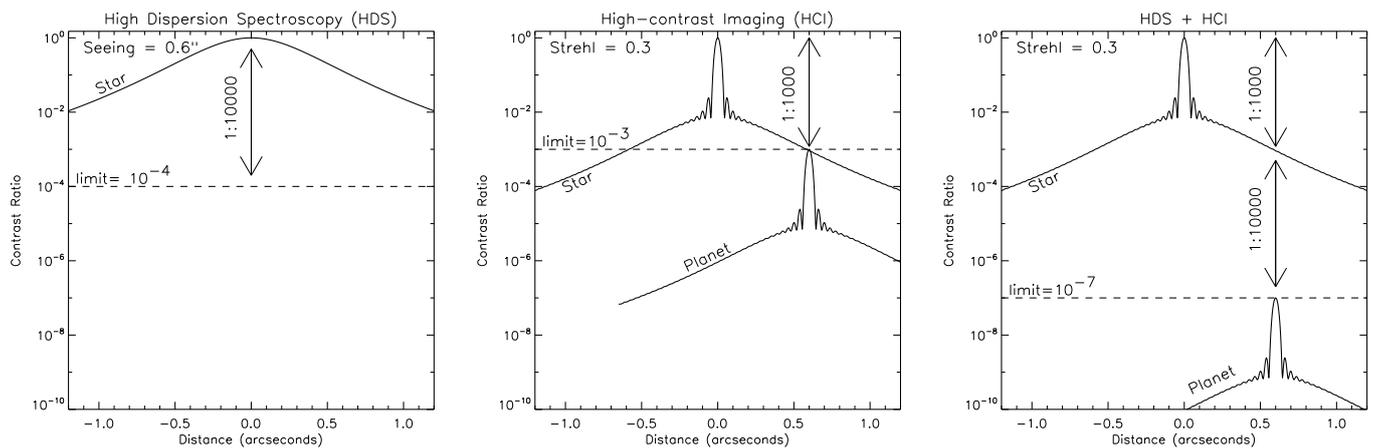}
\caption{Toy model of the HDS+HCI method. The left panel shows the stellar point-spread function (PSF) for conventional, seeing-limited (seeing=0.6 arcseconds) HDS observations, with indicated a contrast of  1$\times 10^{-4}$. This level of contrast has been readily achieved both in the optical  (e.g. \citealp{Leigh03a}) and infrared (e.g. \citealp{Brogi12}), meaning that such planet signal can be detected at a 1:10,000 level in the spectrum of the star. 
 The middle panel shows a model PSF for HCI observations for an adaptive-optics assisted 8m telescope with a Strehl ratio of 0.3 at 0.5 $\mu$m, under the same seeing conditions. The PSF is modeled as the theoretical Airy profile of the telescope combined with a Moffat function as the non-AO-corrected seeing-limited contribution. A hypothetical planet is inserted at an angular distance of 0.6 arcseconds  from the star at a contrast of 1:1,000 with respect to the stellar brightness at the position of the planet. The right panel illustrates that in this example HDS+HCI can achieve a contrast of $10^{-3} \times 10^{-4} = 10^{-7}$ at the planet position.}
\end{figure*}

\citet{Sparks02} were the first to advocate a method that combines coronagraphic imaging with what they called ``spectral deconvolution'', using an integral field spectrograph (IFS; see also \citet{Riaud07}).  The principles laid out in this pioneering work are very similar to what we present here. The technique introduced by \citet{Sparks02} has recently been implemented by \citet{Konopacky13}, who have detected carbon monoxide and water in the spectrum of the directly imaged exoplanet HR8799c. An important aspect of the method described below is the high spectral dispersion. This makes it possible to distinguish between direct starlight and Doppler-shifted starlight reflected off a planet atmosphere. It can also be used to distinguish between the telluric aborption spectrum and Doppler-shifted absorption lines in the thermal spectra of planets very similar to our Earth. \citet{Kawahara14} propose a related technique, combining coronagraphy with HDS to probe hot close-in planets such as hot super-Earths. 

For a planet located at some angular separation from its host star, the starlight at the position of the planet can be reduced up to several orders of magnitude using adaptive optics and coronography. Subsequently, the remaining starlight can be filtered out using high-dispersion spectroscopy, utilizing the significantly different (or Doppler shifted) high-dispersion spectra of the planet and star. While in classical HDS it is the change in the Doppler-shift of the planet signal that is used to separate it from the much stronger (quasi-)stationary stellar and telluric contributions, in the case of HDS+HCI it is the fact that the star-spectrum and telluric absorption are dominant but identical over the whole field, while the planet signal itself is uniquely different and strongly localized at one position in the image. The speckles have the high-dispersion spectrum of the star and are therefore effectively filtered out, influencing only the photon-noise statistics. In contrast to image processing techniques such as ADI and SDI, HDS+HCI will work well at small working angles (a few $\lambda$/D) and is much less sensitive to quasi-static speckle noise that tends to plague HCI. 

While both HDS and HCI can each reach (raw) planet/star contrasts down to $\sim 10^{-5}$, a combination of HDS+HCI can {\sl in principle} reach contrasts of $\sim 10^{-5} \times 10^{-5}$, although in practice this will be limited by photon statistics and/or sky-background. We also note that at a dispersion of R=100,000 the scattered starlight may be coherent, leading to high-frequency fringes, i.e. the speckle pattern as a function of wavelength may not be smooth. This will need to be tested to assure that the above assumptions are correct.

The principle of HDS+HCI is illustrated in Fig. 1, showing in the left panel a toy-model of the stellar point-spread function (PSF) for seeing-limited observations (seeing = 0.6 arcseconds). A hot Jupiter currently can be detected at a contrast of  $<10^{-4}$ in seeing-limited observations,  a level of contrast that has already been achieved both in the optical  (e.g. \citealp{Leigh03a}) and infrared (e.g. \citealp{Brogi12}). The middle panel of Fig. 1 shows the PSF of AO-assisted HCI observations with an 8m telescope at 0.5$\mu$m, with a Strehl ratio of 0.3 under 0.6 arcsecond seeing conditions.  In this example, a raw contrast of $1\times 10^{-3}$ can be achieved 0.6 arcseconds away from the host star (without utilizing data analysis techniques such as SDI or ADI). This means that by combining HDS+HCI, a contrast of $10^{-3} \times 10^{-4} = 10^{-7}$ can be achieved at the planet position, as shown in the right panel of Fig. 1. In addition, each detection of a planet automatically results in a measurement of the radial component of its orbital velocity. By monitoring a system over time, both the planet position and the change in its radial velocity can be used to determine its orbital parameters, including the inclination of the system. 

The contrasts quoted above provide an indication of the currently proven limits of HDS+HCI before systematic effects could start to play a role because both such raw imaging contrasts and HDS signals at these low levels have already been reached. With the starlight reduced by a factor $K$ at the planet position (compared to when no attempt is made to angularly separate the planet from the star), the S/N reached within a certain exposure time will be reduced by a factor $\sqrt{K}$, and for HDS+HCI Eq. 1 becomes
\begin{equation}
\rm{S/N} = \frac{S_{\rm{planet}}}{\sqrt{S_{\rm{star}}/K+\sigma_{\rm{bg}}^2+\sigma_{\rm{RN}}^2+\sigma_{\rm{Dark}}  }} \sqrt{N_{\rm{lines}}},
\end{equation}
meaning that for the same telescope and exposure time, the planet/star contrast improves by a factor $\sqrt{K}$ with respect to classical HDS. In addition, the higher $K$, the higher the relative importance of the sky background ($\sigma_{\rm{bg}}$), which in particular beyond 3 $\mu$m this becomes the dominant noise factor in Eq. 2. It could mean that in the background limited regime existing HCI techniques will be more sensitive than HDS+HCI, depening on the strength and number of planet absorption lines targeted in the observed wavelength regime, and on the relative instrument through-put and bandwidth. In any case, HDS+HCI will have as added value that it also measures the radial component of the orbital velocity of the planet, helping to determine its orbital parameters.

Imaging contrast will be pushed significantly further with the next generation of Extremely Large Telescopes (ELTs). At the diffraction limit, angular resolution scales with $\lambda / D$, where  $D$ is the telescope diameter, meaning that the angular distance in units of diffraction elements increases linearly with $D$. The relevant parameter for these exoplanet observations is the contrast of a point source compared to the local intensities of the star and background. The FWHM of the central part of the  Airy disk scales with $D^{-2}$, while the relative surface brightness of the uncorrected (seeing-limited) part of the PSF is invariant. This means that for the same Strehl ratio, the stellar halo (and the sky background) are reduced by $D^{-2}$, while the collecting power scales as $D^2$. In other words, in the context of a Nyquist-sampled PSF at the diffraction limit of the telescope, the increase in diameter implies that two effects are combined; 1) a modification of the PSF shape for a given Strehl leading to a flux concentration in a PSF core scaled with D$^2$, and 2) a collecting power scaled also as D$^2$. Furthermore, the influence from the star is likely to decrease for larger $D$ because the planet-star separation is larger in terms of $\lambda/D$. Altogether, a certain planet/star contrast can be reached in a time that scales as $1/D^{4}$ or more with the telescope size.

\section{Instrument requirements for HDS+HCI}

Assuming that the positions of the planets are unknown, HDS+HCI requires an adaptive-optics assisted telescope with an Integral Field Spectrograph\footnote{Already such an instrument is planned, called METIS - the Mid-infrared ELT Imager and Spectrograph for the E-ELT \citep{Brandl10}. Its baseline design includes a high-dispersion R=100,000 integral field unit for L (2.9-4.2 $\mu$m) and M (4.5-5.0 $\mu$m) band spectroscopy with an instantaneous wavelength coverage of 0.07$\mu$m, and a field of view of 0.4$\times$1.5 arcseconds cut in 24 slices. To our knowledge, this is the first (planned) instrument that includes a high-dispersion IFS, on a telescope that will utilize adaptive optics techniques. It will be ideally suited for HDS+HCI observations. It may also be equipped with a long-slit unit delivering an instantaneous wavelength coverage of $\sim$0.5 $\mu$m - a powerful addition for planets with a known position.} (IFS). The IFS could either be fiber-fed or use image slicing techniques. In the search for reflected starlight, the optimal spectral dispersion is set by the intrinsic width of the stellar lines, which for slow rotators is dominated by the turbulent velocity fields in the stellar photosphere of  $\sim4$ km sec$^{-1}$. This provides an optimal spectral dispersion of R=75,000 in the optical. The intrinsic widths of absorption lines in a planet thermal spectrum can be significantly more narrow, but integrated over the planet disk are likely to be broadened by the planet's spin velocity. All rocky planets in the Solar System have spin velocities of $<$1 km s$^{-1}$, while those of the gas giants extend up to 12.5 km s$^{-1}$ for Jupiter (and 25 km s$^{-1}$ for $\beta$ Pictoris b; \citealp{Snellen14}). Experience with CRIRES at the VLT shows that a high dispersion is also important for the effective removal of telluric contaminations. It indicates that R=100,000-150,000 is a good compromise. 

The larger the instantaneous spectral coverage, the higher  the enhancement in S/N will be from combining the signals from individual absorption lines, as defined by $N_{\rm{lines}}$ in Eq. 1 \& 2. For probing optical reflected starlight, the 0.4-0.7 $\mu$m wavelength regime, as commonly used by modern optical echelle spectrographs, will cover the densest regions in solar-type spectra. Do note however that for an IFS with R=75,000 this will require a large number of detectors. For cooler host stars, the red part of the optical spectrum is more interesting, up to the 1$\mu$m CCD sensitivity cut-off, where most of the energy of M-dwarfs is emitted, and the molecular bands from titanium oxide and vanadium oxide are prominent. In addition, if Earth-like planets could be probed, the oxygen A-band at 0.76 $\mu$m would be extremely interesting. Similar arguments exist for targeting the intrinsic thermal spectra of exoplanets. Although the entire wavelength range up to 5 $\mu$m is interesting for ground-based HDS (beyond that the sky background is too high), practical limits in wavelength coverage are governed by detector technology and windows of transparency in the Earth atmosphere. The strongest signatures can be expected from water, methane, carbon dioxide, and  also from carbon monoxide for the hottest planets. 

The field of view of the IFS should be at least on the order of $\sim1-2$ arcseconds to include the habitable zones of the nearest solar-type stars. Particularly interesting for optical HDS+HCI are the small inner work angles, meaning that for the ELTs the habitable zones of nearby M-dwarfs at 0.05-0.2 AU could be within reach. This science case allows a significantly smaller field of view. Another important consideration for the instrument design is that typically, image slicer IFUs are not isotropic, meaning that they provide better resolution along the slices than across the slice direction.  Hence, one should observe half of the time with the IFU rotated 90 degrees, and optimally combine the two orientations in software. We note that both the contrast from reflected starlight and thermal emission are expected to decrease significantly as a function of orbital distance. Although young, self-luminous planets can be seen much farther out, after discovery their positions are known and can therefore also be probed with a long-slit HDS to determine their radial velocity and atmospheric constituencies.

\section{Simulations of infrared HDS+HCI observations}

\begin{figure}
\centering
\includegraphics[width=0.5 \textwidth]{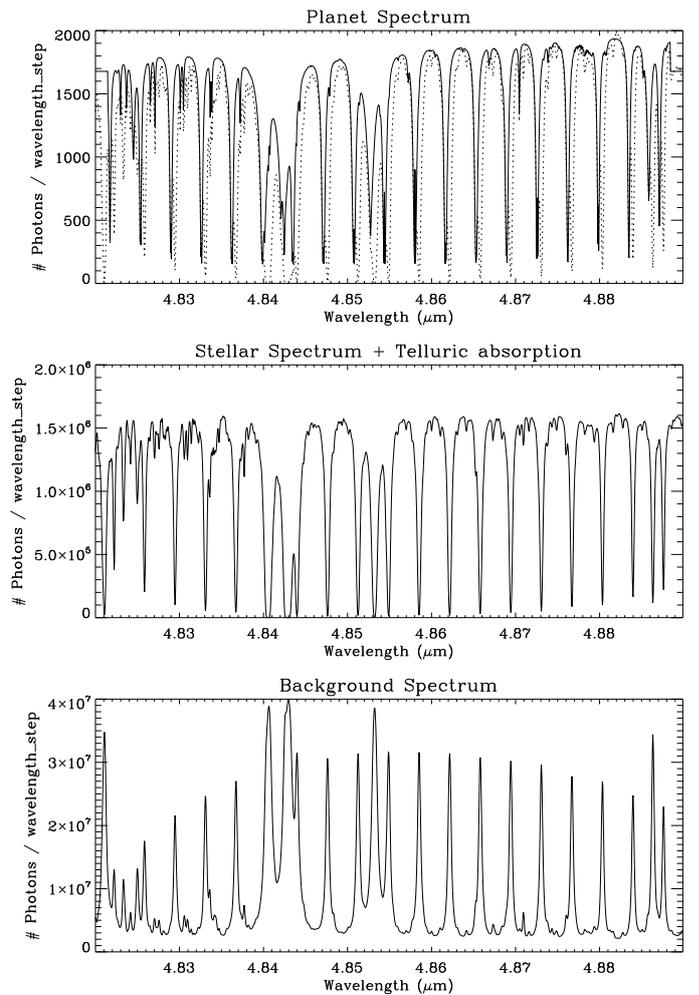}
\caption{Top panel: Simulated E-ELT planet spectrum as photons per $\Delta v = 3$ km s$^{-1}$ wavelength interval. The dotted line indicates the Earth telluric spectrum, which in this case is very similar to the simulated thermal spectrum of the exoplanet (solid line), with the latter shifted by 30 km sec$^{-1}$ in velocity because of its orbital motion around $\alpha$\,Cen\,A. Depending on the time of year, the telluric absorption may show an additional $\pm$30 km sec$^{-1}$ velocity with respect to the exoplanet because of the Earth's orbital velocity around the Sun. Middle panel: The stellar spectrum in photons per wavelength step as seen through the Earth atmosphere, at the sky position of the planet. All the absorption lines are telluric in nature. Bottom panel: The background spectrum in photons per wavelength step per pixel, modeled as contributions from the night sky and emissivity of the E-ELT telescope. The former consists for a large part of bright emission lines at the wavelengths of the telluric absorption lines.}
\end{figure}

\begin{figure}
\centering
\includegraphics[width=0.5 \textwidth]{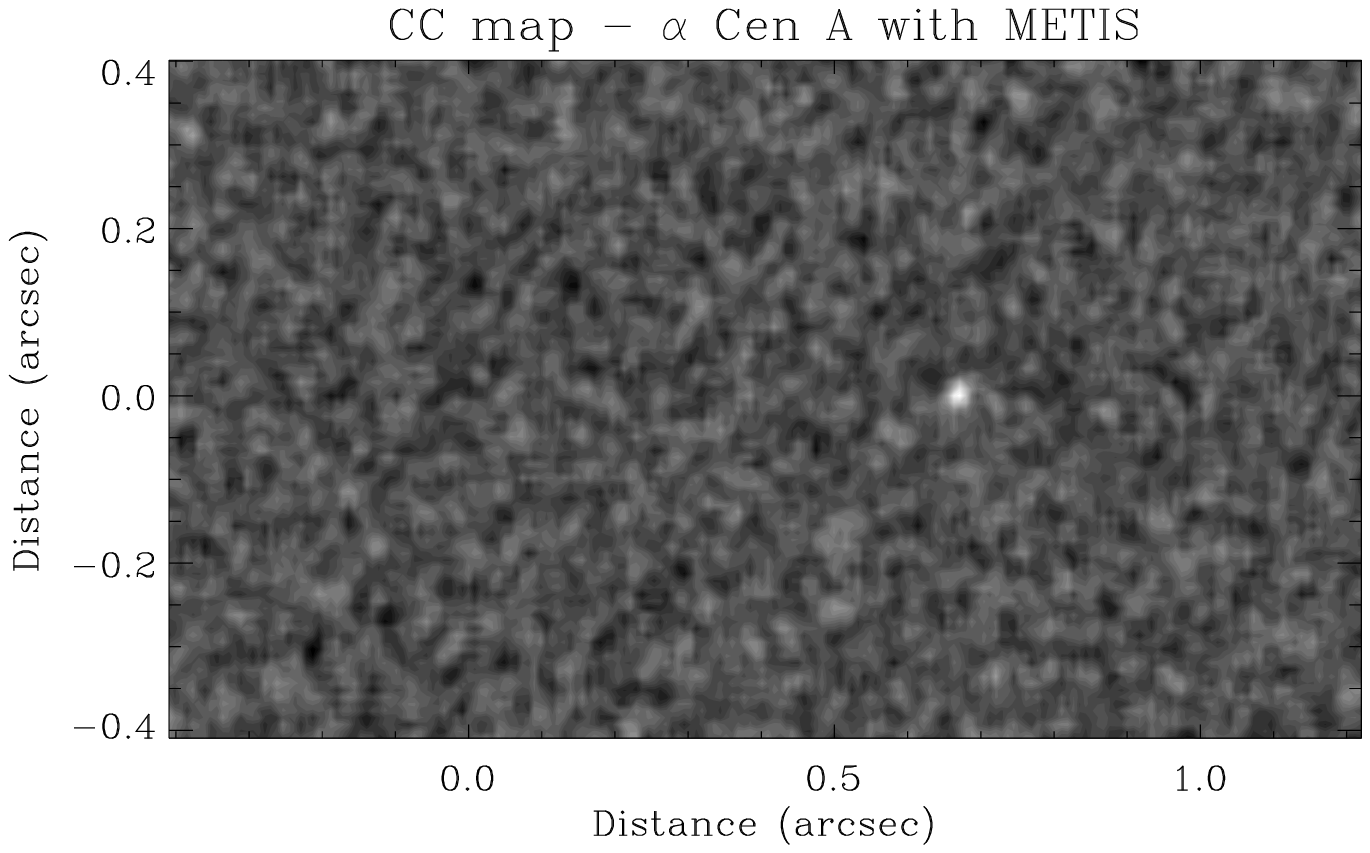}
\includegraphics[width=0.5 \textwidth]{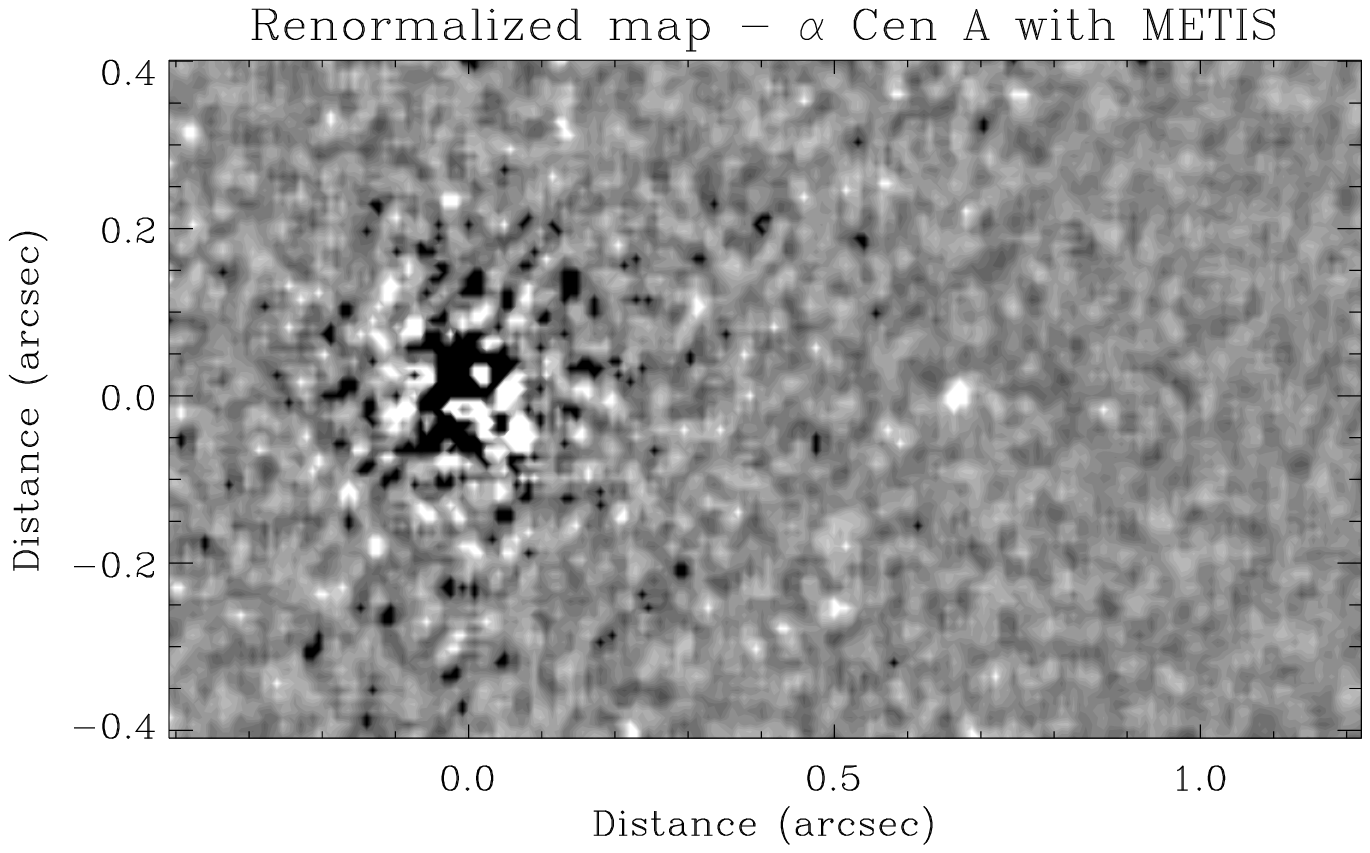}

\caption{Top panel: Slice at the V=30 km sec$^{-1}$ of the cross-correlation data cube showing the simulated planet (R$_{\rm{p}}$= 1.5 R$_{\rm{Earth}}$ and T$_{\rm{eq}}$=300 K) around $\alpha$ Cen A, for a 30 hours exposure with METIS at the E-ELT. The planet is visible 0.7 arcseconds away from the star at an S/N of 8.
Since the cross-correlation technique automatically renormalizes the noise, the star is not visible in this data cube. The bottom panel shows the same but now multiplied by the local contrast to emulate a conventional HCI image.}
\end{figure}

\begin{figure}
\centering
\includegraphics[width=0.5 \textwidth]{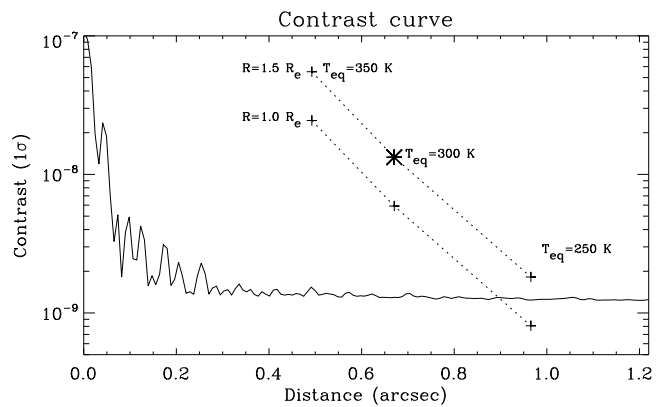}
\caption{The contrast curve for the simulated infrared observations, as presented in Figures 3 and 4, and described in section 4, for an observation of $\alpha$ Cen A with METIS for 30 hours centered at a wavelength of $\lambda = 4.85 \mu$m. The  asterisk signifies the simulated planet detected at an S/N of 8. The dotted lines show scaling laws for planets with the same spectrum, but different effective temperatures (T$_{\rm{eq}}$ = 250, 300, and 350 K) and radii (R = 1.0 and 1.5 R$_{\rm{Earth}}$). No attempt is made to include changes in chemical composition or temperature structure for such planets.}
\end{figure}

\begin{figure}
\centering
\includegraphics[width=0.45 \textwidth]{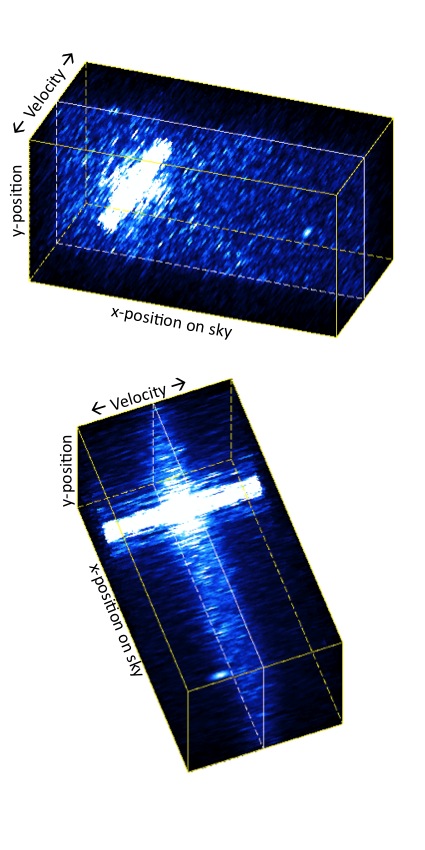}
\caption{Graphical representations of the cross-correlation data cube for METIS observations. All parameters were kept the same as for the simulation presented in section 4. However, for illustrative purposes, the exposure time was increased such to show the planet signal at a S/N of $\sim$20, and  the planet velocity was increased by a factor of 2 to enhance the offset in velocity space. The cross-correlation data cube is renormalized as in the bottom panel of Figure 3. A short movie from this data cube, rotating in sky and velocity angle, is available at http://www.strw.leidenuniv.nl/$\sim$snellen/}

\end{figure}

\begin{table}
\caption{\label{sim1} Main assumptions for the infrared HDS+HCI simulations as presented in Figs. 2 and 3.} 
\begin{tabular}{lr} \hline
Telescope + Instrument\\
Telescope collecting area & 976.3 m$^2$\\
Telescope temperature     & 280 K \\
Telescope emissivity      & 0.15 \\
Telescope+instrument throughput & 15\% \\
AO Strehl (4.85 $\mu$m)   & 0.9 \\ 
Spectral resolution       & R=100,000 \\
Exposure time             & 30 hr\\
Spectral range            & 4.82-4.89 $\mu$m\\
\\
Target: $\alpha$\,Cen\,A  \\
Apparent K magnitude & $-$1.47\\
T$_{\rm{eff}}$ (star)       & 5800 K\\
Stellar radius            & 1.22 R$_{\rm{sun}}$\\
Distance                  & 1.34 pc\\
Planet radius             & 1.5 R$_{\rm{Earth}}$\\
Planet radial velocity    & 30 km sec$^{-1}$\\
T$_{\rm{eff}}$ (planet)       & 300 K\\
Bond albedo                & 0.3\\
Planet spectrum            & Earth-like\\
\hline
\end{tabular}
\end{table}

We simulated infrared HDS+HCI observations with the E-ELT  (see Table \ref{sim1}) of hypothetical planets orbiting $\alpha$\,Cen\,A (K=$-$1.47, T$_{\rm{star}}$=5800 K, R$_{\rm{star}}$=1.22 R$_{\rm{sun}}$, dist= 1.34 pc). The planet radii were varied between 1 and 1.5 R$_{\rm{Earth}}$, with equilibrium temperatures between 255 K (like our Earth) and 355 K. The assumed Bond albedo of A$_{\rm{B}}$=0.3 implies orbital radii in the range  0.63$-$1.22 AU, corresponding to angular distances in the range 0.48$-$0.93 arcsec. The planet's orbital velocity was fixed to 30 km sec$^{-1}$, similar to that of the Earth. The planet spectrum (Fig. 3) was modeled as the thermal spectrum of the Earth scaled by the size of the planet disk, and includes molecular absorption from carbon dioxide, methane, and water vapor. All atmospheric parameters, T/p profile and molecular abundances were fixed to the Earth value. No attempt was made to introduce atmospheric chemistry when changing the atmospheric temperature. The Earth model spectrum was simply scaled according to the change in the blackbody emission at the observed wavelength. 

 We used the baseline instrument parameters of METIS at the E-ELT \citep{Brandl12}. The E-ELT was assumed to have a collecting area of 976.3 m$^2$, at a temperature of 280 Kelvin with an emissivity of 0.15. The total telescope+instrument throughput (excluding atmosphere) was assumed to be 15\%.  We set the wavelength range to 4.82-4.89 $\mu$m, optimizing the combination of strong planet absorption lines and weak telluric absorption. The Earth atmospheric emission and transmission spectra were taken from a recent sky model for ESO exposure time calculators\footnote{See http://www.eso.org/observing/etc/doc/skycalc/helpskycalc.html \#version}. The AO-assisted PSF for METIS has been modeled by \citet{Stuik10}. At this wavelength the extreme AO system on the E-ELT is expected to deliver a Strehl ratio of 0.9 for bright stars. A simulated point-spread function (PSF) for this level of correction, kindly provided by the METIS instrument team, was used for our simulations. The total exposure time varied from a few hours to several nights to determine the S/N limits of these observations.

We first calculated the observed flux from the solar-type star in photons per wavelength interval, and multiplied it with an Earth telluric transmission spectrum. This was subsequently convolved with the model PSF to provide a data cube of stellar flux per spaxel (wavelength interval per IFS pixel). The same was done for the planet, by appropriately scaling down the planet spectrum and Doppler shifting it to the correct wavelength scale, and offsetting the planet position by the angular separation corresponding to its equilibrium temperature. The sky background spectrum was scaled to the collecting area of the E-ELT, the pixel size, and wavelength intervals. This was added to the photon flux expected from the emissivity of the telescope itself. In this way three data cubes were constructed, one with the photon flux from the star, one with the flux from the planet, and one with the flux from the background. Both the background (uniform over the detector) and the contribution from the star (same spectrum over the detector but varying in amplitude) were subsequently assumed to be removed  during the data analysis in such way that only Poisson noise is left from their contributions. We note that this is an idealistic case. Although it has been shown that the HDS method is capable of removing stellar contributions down to $< 10^{-4}$ level, this has not been shown to work down to this level in combination with high contrast imaging, and much remains to be understood for the subtraction of stellar speckles in the high-spectral resolution regime. We used the Gaussian approximation for their noise contributions by computing the square root of these data cubes, which were subsequently multiplied by a random Gaussian noise distribution and added to the planet data cube. The spectrum at each pixel position is subsequently cross-correlated with the planet template spectrum to obtain a three-dimensional data cube, containing x,y sky position, and the cross-correlation signal for velocities between $-$150$\leq V \leq$ +150 km sec$^{-1}$. 

Figure 3 shows a slice of the cross-correlation data cube at the planet velocity (top panel) and a renormalized map (emulating a conventional HCI image - lower panel) for simulations of a planet with R$_{\rm{p}}$= 1.5 R$_{\rm{Earth}}$ and T$_{\rm{eq}}$=300 K for 30 hours of observing time. The planet is clearly visible $\sim$0.7 arcsec away from the star in x-direction, and is detected at an S/N of $\sim$8. In one night (10 hours) it would be detected at an S/N of $\sim$5. Figure 5 shows the contrast achieved as a function of angular distance, with the planet (R$_{\rm{p}}$= 1.5 R$_{\rm{Earth}}$ and T$_{\rm{eq}}$=300 K)  of Fig. 4 indicated by the asterisk. The dotted lines indicate scaling relations for these simulations, for planets with radii of 1.0 and 1.5 R$_{\rm{Earth}}$, and equilibrium temperatures of T$_{\rm{eq}}$=250, 300, and 350 K. Again, these assume the same planet spectra, only scaled in flux by the brightness temperature and planet size, and do not take into account expected changes of chemical composition or atmospheric structure as a function of temperature, and should therefore only be considered as an approximation. An Earth-like planet with T$_{\rm{eq}}$=250 and R$_{\rm{p}}$= 1.5 R$_{\rm{e}}$ has a S/N of $\sim$1 in these 30 hour observations.

Figure 5 shows a graphical representation of the cross correlation data cube for the simulated METIS observations. All parameters are the same as above, except that for illustration purposes the exposure time was increased to more clearly show the planet (now at an S/N of $\sim$20), and the planet velocity was doubled to enhance the offset in velocity space.

If we assume that the targeted planet has already been discovered, e.g. using classical HCI techniques, a long-slit mode can be used to probe the spectrum of the object. This  has the potential advantage of a significantly large instantaneous wavelength coverage, e.g. of 0.5 $\mu$m, instead of 0.07 $\mu$m for METIS. We found that this increases the S/N for a given exposure time by about a factor of $\sim$2. We note however that it may be challenging to center a planet, even when previously imaged, on a 10-20 milliarcsecond wide slit of a single-slit spectrograph (a typical planet around a nearby star could move by this angular distance in one week). The utilization of slit-view cameras or automated centering only works for bright sources, and one would not want to integrate for many hours missing the planet. This issue does not exist for an IFS.

\section{Simulations of optical HDS+HCI observations}

\begin{figure}
\centering
\includegraphics[width=0.5 \textwidth]{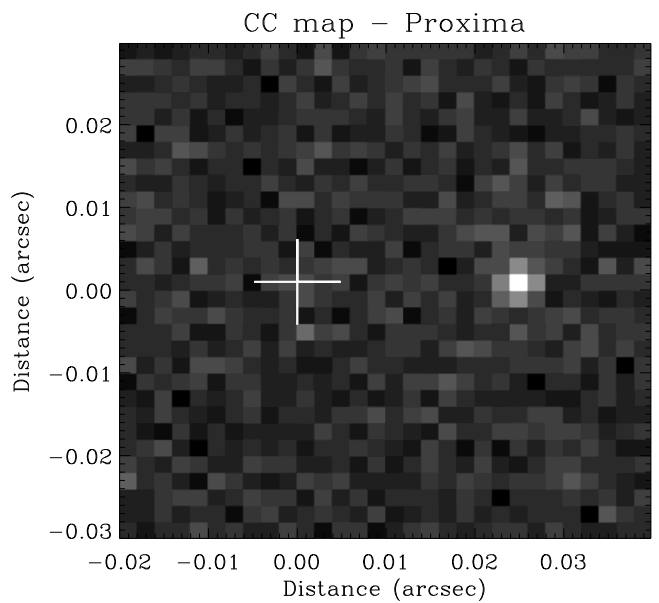}
\caption{HDS+HCI cross-correlation map of 10 hours of optical observations with the E-ELT using a R=100,000 IFS and an adaptive optics system producing a Strehl ratio of 0.3. The hypothetical planet with a radius of R=1.5 R$_{\rm{Earth}}$, albedo of 0.3, and T$_{\rm{eq}}$=280 K such that it is at an orbital radius of 0.032 AU, 25 milliarcseconds from the star. The starlight reflected off the planet is detected at an S/N of $\sim$10.}
\end{figure}
\begin{table}
\caption{Main assumptions for the optical HDS+HCI simulations as presented in Figs. 6.} 
\begin{tabular}{lr} \hline

Telescope + Instrument\\
Telescope collecting area & 976.3 m$^2$\\
Telescope+instrument throughput & 15\% \\
AO Strehl (0.75 $\mu$m)   & 0.3 \\ 
Spectral resolution       & R=100,000 \\
Exposure time             & 10 hr\\
Spectral range            & 0.6-0.9 $\mu$m\\
IFU pixels             & 30$\times$30 2 mas\\
\\
Target: Proxima\,Cen  \\
Apparent V magnitude & $-$11.05\\
T$_{\rm{eff}}$ (star)       & 3040 K\\
Stellar radius            & 0.141 R$_{\rm{sun}}$\\
Distance                  & 1.30 pc\\
Planet radius             & 1.5 R$_{\rm{Earth}}$\\
Planet radial velocity    & 30 km sec$^{-1}$\\
T$_{\rm{eff}}$ (planet)     & 280 K\\
Grey geometric albedo      & 0.3\\
Orbital radius             & 0.032 AU\\
Angular distance from star & 25 mas\\
\hline
\end{tabular}
\end{table}

We also simulated {\sl optical} HDS+HCI observations with the E-ELT, for starlight reflected off planet atmospheres. Although currently the optical high-contrast imaging instrument for the E-ELT, EPICS (\citealp{Kasper10}), is proposed with a low-dispersion IFS, we would like to make a case for a high-dispersion IFS.
At these wavelengths the angular resolving power of the E-ELT will be well suited to target planets in the habitable zones of nearby M-dwarfs. It is clear that an extreme adaptive optics system for the optical wavelength regime will be technically challenging (\citealp{Kasper10}), and we therefore assume a lower Strehl ratio of 0.3 under 0.6 arcsec seeing conditions. For the integral field spectrograph we assume a spectral dispersion of R=100,000 with 30$\times$30 elements of 2.0 milliarcsecond pixels. Hence the field of view is 60$\times$60 mas, well matched to resolving the habitable zones of nearby M-dwarfs. The wavelength range is set to 0.6$-$0.9 $\mu$m. Extending this to shorter wavelengths does not significantly increase the sensitivity because of the red colors of the host stars. Also, we note that this set-up will require several tens of CCDs.

In our simulations we target our nearest neighbor Proxima Centauri (V=11.05 mag, T$_{\rm{star}}$=3040 K, R$_{\rm{star}}$=0.141 R$_{\rm{sun}}$, Dist= 1.30 pc). We assume a hypothetical planet with a radius of R=1.5 R$_{\rm{Earth}}$, a gray albedo of 0.3, and T$_{\rm{eq}}$=280 K such that it is at an orbital radius of 0.032 AU, 25 milliarcseconds from the star. We assume that half of the planet is seen to be illuminated by its host star, as expected at the greatest elongation. 

We subsequently follow a similar strategy as for the METIS infrared simulations above. Again, we assume a telescope diameter of 39m, with an effective collecting area of 976 m$^2$, and a total  telescope+instrument throughput of 15\%.  Since we do not have a detailed simulation of the telescope PSF at optical wavelength, we approximate the theoretical, monochromatic PSF of the E-ELT at optical wavelengths by simply scaling the angular size of the diffraction-limited PSF at L-band.  We do not consider a polychromatic PSF because this would increase the computational time by an order of magnitude, and would only result in a low-frequency modulation of the observed spectra per pixel, which can easily be filtered out. 
The effect of seeing has been taken into account by scaling the amplitude of the PSF to 30\% according to the Strehl ratio, and adding a seeing limited halo with a relative strength of 70\%.  This halo was modeled as a Moffatt profile\footnote{Moffat profile is $F(r)=F_0 [1+(r/\alpha)^2]^{-\beta}$} with $\beta = 2.5$ and $\alpha$ = 265.4, to match a seeing FHWM of 0.6 arcseconds.

As above, we constructed a three-dimensional data cube, with x-, and y-position, and wavelength on its axes. The spectrum of the star is calculated, assuming 10 hours of observing time with the E-ELT, in photons per wavelength step, and translated to photons per wavelength step per pixel using the normalized PSF. The same procedure is followed for the planet signal, except that it is first scaled according to the planet-star contrast (in this case 6.0$\times 10^{-7}$), shifted in velocity by +30 km sec$^{-1}$, and added to the data cube at the calculated planet position. Since we assume that the starlight can be reduced down to the photon noise, to emulate noise only the square-root of the number of photons per pixel for the star is added to each pixel, and multiplied by a random selection from a Gaussian distribution with $\sigma = 1$. 
As in the case of infrared HDS+HCI, we assume an idealistic case. Despite much success with HDS alone, the ability to remove stellar contributions to $<10^{-4}$ when combining with HCI remains to be proven because of the unknown behavior of stellar speckles at high-spectral resolution.
 Although some regions in the optical wavelength regime are affected by telluric absorption, for simplicity we ignore this for these optical simulations. Including it should not make a significant difference to our S/N calculations. The sky background is also ignored, since  it is more than 23 magnitudes fainter ($<5 \times 10^{-10}$) per pixel than the star at a good observing site. 

Each pixel location is subsequently cross-correlated with the stellar spectrum to produce a cross-correlation data cube, with x,y position and velocity (-150 to +150 km sec$^{-1}$) on its axes. The slice at the planet velocity is shown in Figure 7. Since the cross-correlation technique automatically renormalizes the noise, the star is not visible in this data cube. The planet, at a contrast level of 6.0$\times 10^{-7}$ at 25 milli-arcseconds from the star, is detected at S/N$\sim$10.

\section{Summary and outlook}

We have conducted simulations of HDS+HCI observations with the E-ELT, in the infrared and optical wavelength regime. The infrared observations target absorption lines in the thermal emission spectrum of the planet. The METIS instrument will have the high-dispersion (R=100,000) IFS capabilities to perform such observations, and we show that at M-band, rocky planets in the habitable zone of $\alpha$ Centauri A are within the realm of this instrument.  Since the angular separation of the habitable zone for this star is quite far out, it is expected that such planet will be in the sky background limited regime for the E-ELT. It may well be that  more classical high-contrast imaging techniques, making use of angular and/or spectral differential imaging (ADI \& SDI) will be similar, or even more powerful than HDS+HCI under these conditions since they make a more efficient use of the available photons. For more distant stars the relative contribution of the sky background is even larger, but the planets will  be closer in in angular space, towards the regime where ADI and SDI are less effective. In addition, HDS+HCI observation will provide the radial component of the planet orbital velocity, line strength of molecular absorption in the planet atmosphere, and can be used to determine the planet spin (see below). 

Optical HDS+HCI is technically more challenging, because of the more stringent requirements on the telescope adaptive optics, but it will be more powerful than infrared observations because the sky background plays a significantly smaller role. We show that  with a  Strehl ratio of 0.3, one night of observation of Proxima Centauri with the E-ELT could reveal a rocky planet in its habitable zone, with a few dozen other late M-dwarf that could be surveyed in this way. 

The discovery and characterization of the planetary systems around our stellar neighbors constitute fundamental measurements of great scientific and philosophical value. The HDS+HCI technique has the potential to detect and characterize temperate rocky planets in their habitable zones. Exoplanet scientists should therefore not shy away from claiming a large fraction of the time on the future ELTs to make such observations.

\subsection*{First HDS+HCI observations}

HDS+HCI observations of unknown planetary systems will require a high-dispersion (R$\approx 100,000$) integral field spectrograph with adaptive optics capabilities. To our knowledge such a system does not exist on current telescopes. However, if the position of the planet is known, then a slit spectrograph can also be used, orientated in such way that it encompasses both the planet and the host star. We tested the HDS+HCI technique on the well-known young planetary system $\beta$ Pictoris b, using the R=100,000 CRIRES spectrograph on the VLT, which has been presented in a separate publication \citep{Snellen14}. Although the observations were conducted under mediocre (1-1.3 arcsec) seeing conditions, the starlight at the planet position was reduced by a factor of 8 to 30 compared to the peak of the stellar profile using the MACAO adaptive optics system. The 2.3 $\mu$m bandhead of carbon monoxide was targeted, and the planet, $\sim4000$ times fainter than the star at this wavelength, was detected with an S/N of 6.4 in only one hour of observations including overheads. For comparison, $\tau$ Bo\"otis b, at a similar planet/star contrast and with the star of the same brightness as $\beta$ Pictoris, was detected at a similar significance with classical HDS \citep{Brogi12} but in 18 hours of observing time. The factor of $\sim$20 gain in observing time is as expected from the factor of 8-30 reduction in starlight at the planet position using HDS+HCI. 

As a testemony to the potential power of the HDS+HCI technique, the observations revealed the planet orbital velocity, and its spin - a first for an extrasolar planet. 

\begin{acknowledgements}
This work is part of the research programmes PEPSci and VICI 639.043.107, which are financed by the Netherlands Organisation for Scientific Research (NWO). JLB acknowledges funding from a NASA Sagan Fellowship. MB acknowledges funding from a NASA Hubble Fellowship.
\end{acknowledgements}


\bibliographystyle{aa}
\bibliography{bibtex_jul13a}

\end{document}